\documentclass[aps,pra,twocolumn,superscriptaddress]{revtex4-1}
\usepackage{bm}
\usepackage{graphicx}
\usepackage{color}
\usepackage{braket}
\usepackage{amsmath,amssymb,amsfonts,amsthm,mathtools}
\usepackage{enumerate}
\usepackage{enumitem}
\usepackage[colorlinks=true,linkcolor=blue,anchorcolor=red,citecolor=blue,urlcolor=blue]{hyperref}
\usepackage{titlesec}
\usepackage{tikz-cd}
\usepackage{float}
\usepackage{multirow}
\usepackage[title]{appendix}
\usepackage{pdfpages}
\usepackage{changes}
\usepackage{stackengine}
\usepackage{makecell}
\usepackage{caption}
\usepackage{pdfpages}

\makeatletter
\AtBeginDocument{\let\LS@rot\@undefined}
\makeatother

\makeatletter 
\renewcommand\NAT@biblabelnum[1]{#1.} 
\makeatother

\begin{document}
\author{Zheng Zhang}
\email{zhengzhang@lut.edu.cn}
\affiliation{Department of Physics, School of Science, Lanzhou University of Technology, Lanzhou 730050, China}
\affiliation{Department of Physics, The University of Hong Kong, Pokfulam Road, Hong Kong, China}

\title{Classical Resolution of the Gibbs Paradox from the Equal Probability Principle: An Informational Perspective }

\begin{abstract}
The Gibbs paradox is a conventional paradox in classical statistical mechanics, typically resolved by invoking quantum indistinguishability through the $1/N!$ correction. In this letter, we present a resolution within classical ensemble theory, which relies solely on the equal probability principle and does not invoke the $1/N!$ correction. Our resolution can be naturally interpretated from a purely informational perspective, where the Gibbs entropy is explicitly regarded as the Shannon entropy, quantifying ignorance rather than disorder. From this informational perspective, we also clarify the connection between information and extractable work in the gas mixing processes. Our work opens a new avenue to reconsider the role of information in statistical mechanics.
\end{abstract}

\maketitle

\section{Introduction}
\label{sec1}

The Gibbs paradox is a traditional paradox in classical statistical mechanics \cite{Gibbs1874,Gibbs1902,Kampen1984,Jaynes1992,Darrigol2018}. When classical ensemble theory is applied to an ideal gas, it yields a non-extensive entropy. This becomes particularly apparent when considering the mixing of identical gases, where the non-extensivity leads to paradoxical consequences. Traditionally, this paradox is resolved by involving quantum mechanics, which introduces a $1/N!$ correction factor due to the particle indistinguishablity \cite{Plank1925,Huang1963,Feynman1972,Zubarev1974,Landau1975,Toda1978,Callen1985}. This correction restores the extensivity of entropy and eliminates the paradox. However, in many regimes where statistic mechanics remains applicable, quantum effects are ignorable, even particles are experimentally distinguishable \cite{Frenkel2014}. Many researches attempted to resolve the Gibbs paradox within the framework of classical statistics \cite{Kampen1984,Jaynes1992,Tseng2002,Swendsen2002,Swendsen2018,Cheng2009,Versteegh2011,Dieks2018,Peters2014,Murashita2017,Darrigol2018,Saunders2018,Sasa2022,Lairez2024}, and it is now a common sense that the paradox admits a classical resolution, although the approaches adopted in these studies are diverse. 

In this letter, we present a new classical resolution to the Gibbs paradox. Compared to previous works, our resolution is fundamentally distinct by two key features. First, unlike previous works that resort to complicated arguments or subtle principles, our resolution relies solely on the equal probability principle of classical ensemble theory and does not require the $1/N!$ correction, which makes our resolution rather straightforward and simple. Contract to traditional belief, we demonstrate that non-extensivity of the entropy calculated by statistical mechanics causes no paradox if the equal probability principle is strictly maintained.

Another fundamental distinction is that we adopt a purely informational perspective to understand the paradox. This perspective is motivated by the form of the Gibbs entropy in classical ensemble theory: 
\begin{equation}\label{en1}
S=-k\int \rho(p,q) \ln \rho(p,q) dpdq,
\end{equation}
where $\rho(p,q)$ is the probability density distribution in phase space. This expression is formally identical to Shannon's definition of information entropy \cite{Shannon1948}. E. T. Jaynes first took this coincidence seriously and proposed that the Gibbs entropy should be understood as a kind of information entropy, which measures our ignorance about the microscopic state of the system, rather than its disorder. We show that the informational interpretation of entropy offers a very natural understanding for our resolution. Moreover, the informational perspective helps us revealing the important role of information in the Gibbs paradox. We find that whether there is entropy change in the gas mixing process depends on whether there is information loss in the process.  In recent years, the relation between information and work has been explored in the study of information thermodynamics \cite{parrondo2015,Lloyd1989,Sagawa2008,Sagawa2009,Sagawa2010,Horowitz2010,Hasegawa2010,Takara2010,Horowitz2011,Esposito2011,Sagawa2012,Deffner2012,Horowitz2014}. Using the information-work relation, we can reach a deeper understanding of the Gibbs paradox. We believe the informational perspective we present is not limited to the Gibbs paradox, but signifies a paradigm shift in statistical mechanics, in which basic concepts such as entropy and information will be rethought.

\section{Review of  the Gibbs paradox}
\subsection{The Gibbs paradox}
Let us first briefly review the origin of the Gibbs paradox in classical statistical mechanics. 
The foundation of classical ensemble theory is the equal probability principle, which says for a microcanonical ensemble (fixed $N,V,E$), all accessible phase space points on the energy surface $H(p,q) = E$ are equally probable, with $\rho(p,q) = \text{constant}$. From this principle, we can derive the probability density distribution for canonical ensembles (fixed $N,V,T$):
\begin{equation}\label{cannon}
\rho(p,q)=\frac{e^{-\beta H(p,q)}}{Z},
\end{equation}
with 
\begin{equation}\label{Z}
Z=\int e^{-\beta H(p,q)} dpdq, 
\end{equation}
being the partition function, where $\beta=1/kT$. Once the partition function is obtained, various quantities in statistical mechanics can be calculated. 
For canonical ensembles, the Gibbs entropy can be calculated by plugging Eq. (\ref{cannon}) into Eq. (\ref{en1}), or by using
\begin{equation}\label{sz}
S=k(\ln Z-\beta \frac{\partial }{\partial \beta}\ln Z),
\end{equation}
where $Z$ is given by Eq. (\ref{Z}).

Applying Eq. (\ref{Z}) and Eq. (\ref{sz}) to calculate the Gibbs entropy for an ideal gas with volume $V$, temperature $T$, and particle number $N$, one obtains a non-extensive entropy:
\begin{equation}\label{en2}
S_{\mathrm{id}}(N,V,T)=\frac{3}{2}Nk[1+\ln (2\pi mkT)]+Nk\ln V.
\end{equation}

The traditional view holds that the non-extensive entropy $S_{\mathrm{id}}(N,V,T)$ causes the Gibbs paradox. Consider a box filled with an ideal gas, partitioned by a wall into two halves A and B, which contains $N_A$ and $N_B$ particles respectively. For simplicity, we assume $N_A=N_B=N$. All particles have the same mass $m$. When the system is in equilibrium, both sides occupy volume $V$. If we assume the entropy is additive, then the total entropy of the system is:
\begin{equation}\label{S1}
S_{\mathrm{t}}^{(1)}=S_A+S_B=2S_{\mathrm{id}}(N,V,T).
\end{equation}
Then we remove the wall, the system will equilibrate to a new state with $2N$ particles in volume $2V$, having entropy
\begin{equation}\label{S2}
S_{\mathrm{t}}^{(2)}=S_{\mathrm{id}}(2N,2V,T).
\end{equation}
This results in an entropy increase:
\begin{equation}\label{dS}
\Delta S = S_{\mathrm{t}}^{(2)}-S_{\mathrm{t}}^{(1)}=2Nk\ln 2.
\end{equation}
When the gases in the two compartments are of different types, $\Delta S$ can be interpreted as the ``mixing entropy'' due to the irreversible mixing of distinct substances. However, if the gases are of the same type, the entropy increase appears unphysical—it contradicts our intuition that no thermodynamic change should occur upon removing a partition between two identical systems. This discrepancy constitutes the famous Gibbs paradox.

\subsection{Traditional resolution}
In the conventional view, the Gibbs paradox arises from the failure of the classical ensemble theory to predict the correct thermodynamic entropy (the entropy related to experimental measurements), i.e., there is a discrepancy between $S_{\mathrm{id}}(N,V,T)$ predicted by the classical ensemble theory and the thermodynamic entropy $S_{\mathrm{id}}^{\mathrm{th}}(N,V,T)$ for an ideal gas. In general, $S_{\mathrm{id}}(N,V,T)$ and $S_{\mathrm{id}}^{\mathrm{th}}(N,V,T)$ can differ by a function $f(N)$ \cite{Jaynes1992}, i.e., $S_{\mathrm{id}}^{\mathrm{th}}(N,V,T)=S_{\mathrm{id}}(N,V,T)+f(N)$. To  correctly predict the entropy change in the gas mixing process, it is thought that $f(N)=Nf(1)-N\ln N$\cite{Jaynes1992,Murashita2017}, or more precisely, $f(N)=Nf(1)-\ln N!$ \cite{Murashita2017}. In practical calculations, to obtain an entropy which is consistent with the thermodynamic entropy, people often introduce a $1/N!$ factor to the partition function in Eq. (\ref{Z}):
\begin{equation}
Z=\frac{1}{N!}\int e^{-\beta H(p,q)} dpdq, 
\end{equation}
With this correction, Eq. (\ref{sz}) can give an entropy which is thought to be consistent with the thermodynamic entropy for an ideal gas. The introduction of the $1/N!$ factor is usually attributed to the particle indistinguishability in quantum mechanics \cite{Plank1925,Huang1963,Feynman1972,Zubarev1974,Landau1975,Toda1978,Callen1985}.  However, for classical systems in which quantum effects are ignorable, this explanation is questionable. Recent studies try to offer a explanation for the $1/N!$ factor within classical statistical mechanics \cite{Tseng2002,Swendsen2002,Swendsen2018,Cheng2009,Versteegh2011,Dieks2018,Darrigol2018,Saunders2018,Sasa2022,Lairez2024}, but a concise explanation is lacking.

\section{Resolution based on the equal probability principle}
In contrast to traditional resolutions of the Gibbs paradox, we think that the classical ensemble theory is sufficient to predict the correct thermodynamic entropy, i.e., $S_{\mathrm{id}}(N,V,T)$ in Eq. (\ref{en2}) can be identified as the thermodynamic entropy $S_{\mathrm{id}}^{\mathrm{th}}(N,V,T)$ without causing any paradox. As a result, the traditional $1/N!$ correction to the partition function in Eq. (\ref{Z}) is not needed. In our view, the Gibbs paradox arises from a naive application of the additivity of the Gibbs entropy in Eq. (\ref{S1}) when calculating the total entropy for a composite system. In fact, as we will show, the additivity of the Gibbs entropy conflicts with Eqs. (\ref{cannon})-(\ref{sz}), which are directly obtained from the equal probability principle. Thus, to resolve the Gibbs paradox within the classical ensemble theory, what we need is to calculate the total entropy directly from the basic rules of the theory, without referring to any other postulations such as the additivity of entropy. 


Let us begin with analyzing the gas mixing process involving two different types of gases, which we label as type A (left side) and type B (right side). We denote the coordinates and momenta of the $N$ particles of type A as $\vec{q}_1, \vec{p}_1, \vec{q}_2, \vec{p}_2,..., \vec{q}_N, \vec{p}_N$, and those of the $N$ particles of type B as $\vec{q}_{N+1}, \vec{p}_{N+1}, \vec{q}_{N+2}, \vec{p}_{N+2},..., \vec{q}_{2N}, \vec{p}_{2N}$. The total phase space of the system is $6N$ dimensional. 

Before removing the wall, we \textit{know} that all type-A particles are confined to the left half, and all type-B particles to the right half. So the coordinates are constrained as $0<q_{i,x}< L$ for $i=1,2,...,N$, and $L<q_{i,x}<2L$ for $i=N+1,N+2,...,2N$, where $2L$ is the length of the box along the $x$ direction, i.e., the direction perpendicular to the wall. The partition function of the whole system is
\begin{equation}\label{Z1}
\begin{aligned}
Z=& \int dq_{1,y}dq_{1,z} \cdot\cdot\cdot dq_{2N,y}dq_{2N ,z}\\ &\int_{0}^{L} dq_{1,x} 
\cdot \cdot\cdot \int_{0}^{L} dq_{N,x} \int_{L}^{2L}dq_{N+1,x}\cdot \cdot\cdot \int_{L}^{2L}dq_{2N,x} \\
&\int d\vec{p}_1\cdot\cdot\cdot d\vec{p}_{2N}\ e^{-\beta \sum_i^{2N} \frac{\vec{p}_i^2}{2m}}= Z_A\cdot Z_B,
\end{aligned}
\end{equation}
where $Z_A$ and $Z_B$ are partition functions for subsystem A and B respectively, which are given as
\begin{equation}
\begin{aligned}
Z_A=&\int dq_{1,y}dq_{1,z}\cdot\cdot\cdot dq_{N,y}dq_{N,z}\int_0^L dq_{1,x} \cdot\cdot\cdot\int_0^L dq_{N,x}\\
&\int d\vec{p}_1\cdot \cdot\cdot d\vec{p}_N e^{-\beta \sum_i^{N} \frac{\vec{p}_i^2}{2m}}= Z_B.
\end{aligned}
\end{equation}
Evaluating Eq. (\ref{Z1}) and applying Eq. (\ref{sz}), we obtain the total entropy before mixing:
\begin{equation}\label{S12}
S_{\mathrm{t}}^{(1)}=S_A+S_B=2S_{\mathrm{id}}(N,V,T),
\end{equation}
which is identical to Eq. (\ref{S1}). After mixing, the total entropy $S^{(2)}_t$ is still given by Eq. (\ref{S2}). Thus, the entropy change is $2Nk\ln2$ as in Eq. (\ref{dS}), which can be interpreted as the “mixing entropy” between different types of gas.


Now, we turn to analyze the case where both sides contain the same type of gas. In this case, we denote the $2N$ particles' coordinates and momenta as $\vec{q}_1, \vec{p}_1, \vec{q}_2, \vec{p}_2,..., \vec{q}_{2N}, \vec{p}_{2N}$. Here we note in classical statistics, particles are distinguishable, and the labels $1,2,...,N$ can be regarded as resigned according to their distinguishable properties (although in our condition we do not know these information). 

Before removing the wall, since we \textit{do not know} the side of each particle, the phase space should be made to contain all possible configurations of particle partition. Each possibility can be represented by
\begin{equation}
\begin{aligned}
&0<q_{\sigma(i),x}<L, \ \mathrm{for} \ i=1,2,...,N;\\
&L<q_{\sigma(i),x}<2L, \ \mathrm{for} \ i=N+1,N+2,...,2N,\\
\end{aligned}
\end{equation}
where $\sigma(i)$ is a permutation of the indices $1,2,...,2N$. All possible configurations of particle partition are given by running over all permutations and drop repeated cases. The total number of possibilities are $(2N)!/(N!)^2$, and accordingly, the phase space is split into $(2N)!/(N!)^2$ disconnected regions. The equal probability principle requires us to assign equal probabilities for phase points with equal energy in all these disconnected regions. Thus the total partition function becomes:
\begin{equation}
\begin{aligned}
Z=& \sum_{\sigma}'\int dq_{1,y}dq_{1,z} \cdot\cdot\cdot dq_{2N,y}dq_{2N ,z}\int_{0}^{L} dq_{\sigma(1),x} \cdot\cdot\cdot\\& \int_{0}^{L} dq_{\sigma(N),x}\int_{L}^{2L}dq_{\sigma(N+1),x}\cdot \cdot\cdot \int_{L}^{2L}dq_{\sigma(2N),x} \\
&\int d\vec{p}_1\cdot\cdot\cdot d\vec{p}_{2N}\ e^{-\beta \sum_i^{2N} \frac{\vec{p}_i^2}{2m}}= \frac{(2N)!}{(N!)^2} Z_0,
\end{aligned}
\end{equation}
where $\sum'$ means sum over non-repeated configurations of particle partition under permutations, and $Z_0$ takes the value in Eq. (\ref{Z1}). Then the total entropy is given by (using the sterling formula $\ln N!\simeq N\ln N-N$):
\begin{equation}\label{S11}
S_{\mathrm{t}}^{(1)}=2S_{\mathrm{id}}(N,V,T)+2Nk\ln 2,
\end{equation}
Comparing Eq. (\ref{S11}) with Eq. (\ref{S12}), there is an additional term $2Nk\ln 2$, arising from the sum over all disconnected regions in phase space. This term destroys the additivity of entropy and demonstrates that additivity conflicts with Eqs. (\ref{cannon})-(\ref{sz}). After mixing, the total entropy $S^{(2)}_t$ is given by Eq. (\ref{S2}). Then the entropy change is $\Delta S=S_{\mathrm{t}}^{(2)}-S_{\mathrm{t}}^{(1)}=0$, which is consistent with the thermodynamic entropy change. Therefore, the Gibbs paradox is resolved.

Since the entropy change calculated from classical ensemble theory coincides with the thermodynamic entropy change, it is permissible for us to identify $S_{\mathrm{id}}(N,V,T)$ in Eq. (\ref{en2}) with the thermodynamic entropy $S_{\mathrm{id}}^{\mathrm{th}}(N,V,T)$ without causing any paradox.

\section{Informational perspective interpretation}
Although the Gibbs paradox is resolved within the classical ensemble theory, its physical meaning still requires interpretation. Interestingly, we find that the informational perspective of entropy \cite{Jaynes1957}, though not widely acknowledged, provides a suitable interpretation for our resolution. 

Taking the similarity between the Gibbs entropy in Eq. (\ref{en1}) and information entropy seriously, the informational perspective regards the Gibbs entropy as a kind of information entropy, which measures our ignorance about the system rather than the disorder. 
In this view, the entropy is not an intrinsic property of the system, but depends on our knowledge: the more we know about the system, the lower its entropy. 
We now show how this purely informational view provides a suitable interpretation for our resolution of the Gibbs paradox. 

Consider $2N$ particles equally partitioned into subsystem A and B. Our ignorance for the system can be decomposed into two independent parts: (1) Ignorance about how the particles are divided between the two subsystems (i.e., which particles belong to which part). (2) Ignorance about the kinetic states of subsystem A and subsystem B. The total entropy is then 
\begin{equation}\label{deco}
S_{\mathrm{t}}=S_{\mathrm{d}}+S_A+S_B,
\end{equation}
where $S_{\mathrm{d}}$ is contributed by the ignorance of the partitioning details, and $S_A, S_B$ measure ignorance of the kinetic state of A and B respectively. We note that this decomposition Eq. (\ref{deco}) can be  derived directly from the composition law of information entropy (see Appendix). The composition law explains why the additivity is not valid for the Gibbs entropy in general.

When the gases in A and B are of different types, there is no ignorance about the partition (we know exactly which particles belong to which subsystem), so $S_{\mathrm{d}}=0$. The total entropy is simply $S_A+S_B$. If the wall is removed, each particle's coordinate can occupy more values, increasing our ignorance about the system's kinetic state and thus the entropy. This explains the nonzero mixing entropy in Eq. (\ref{dS}).

When the gases in A and B are of the same type, there are $W_{\mathrm{d}}=(2N)!/(N!)^2$ ways to equally partition the $2N$ particles. Since we have no preference for any way, our ignorance about the partition is measured by the entropy $S_{\mathrm{d}}=k\ln W_{\mathrm{d}}\simeq 2Nk\ln2$. This explains the additional term in Eq. (\ref{S11}) from summing over disconnected regions in phase space. If the wall is removed, the ignorance about partition vanishes, but the ignorance about kinetic state increases, resulting a zero change of the entropy. To understand the zero change of entropy intuitively, we can view the ignorance about partition as ignorance about coordinates, since the partition constrain the coordinates. After wall removal, we are neither more nor less ignorant about the coordinate for each particle, so the entropy is not changed.

\begin{figure}
\centering
\includegraphics[width=0.85\linewidth]{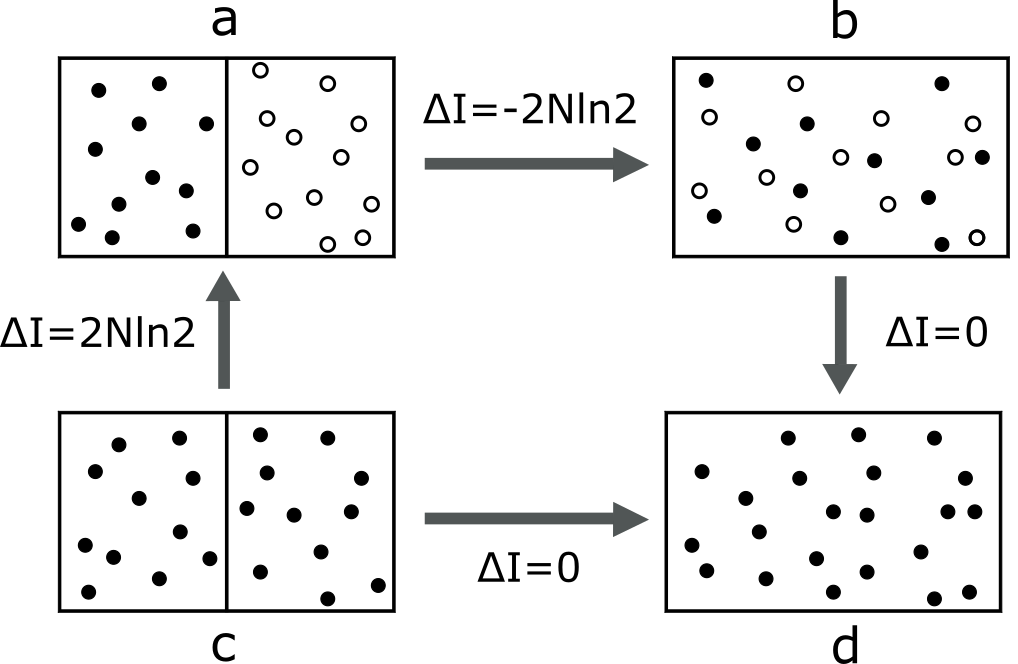} 
\caption{Information differences among four scenarios in the Gibbs paradox. Colored circles means different types of molecules. Compared to scenario (c), scenario (a) has additional $2N\ln2$ nats information regarding the side to which each particle belongs. From (a) to (b), the wall is removed, the $2N\ln2$ nats information is lost, resulting a mixing entropy $2kN\ln2$. Scenarios (b)(c) and (d) has no information difference (or entropy difference), since we are equally ignorant about the positions of particles in these scenarios. }
\label{fig1}
\end{figure}

Information, defined as negative entropy, represents a reduction in our ignorance about a system. The entropy differences for the four scenarios we discussed can be naturally understood in terms of information gain or loss (see Fig.~\ref{fig1}). For instance, before removing the wall, the entropy of the system containing two different types of gases is lower than that of the system with identical gases by an amount of $2Nk\ln2$. This difference corresponds to $I=2N\ln 2$ nats information (note that thermodynamic entropy and information entropy differ by a Boltzmann constant $k$), stemming from the knowledge of which side each particle resides. After removing the wall, this information is lost, resulting in an increase in entropy by $2Nk\ln2$.

\section{Information and work}
One main insight from the study of thermodynamics of information is that information can be used to do work \cite{parrondo2015,Sagawa2008,Horowitz2010,Sagawa2012}. This insight suggests that information is not merely a subjective quantity, but has objective physical consequences. In the gas mixing process, information can also be used to extract work from the system. When the types of gases are different, the $I=2N\ln2$ nats of partition information can be harnessed by replacing the wall with a movable semi-permeable membrane, which selectively permits one gas type to pass. Over time, particle accumulation on one side displaces the membrane, enabling work extraction. Eventually, all particles end up on one side, and the membrane reaches the boundary. The final state corresponds to the fully mixed state, which has a higher entropy than the initial state by $2kN\ln 2$. Throughout this process, the partition information in the initial state is lost, but we extract work from the system (note that without the partition information, controlled work extraction would be impossible).

In generally, for canonical systems, the work can be extracted is constrained by the second law of thermodynamics:
\begin{equation}\label{WF}
W \leq -\Delta F,
\end{equation}
where $\Delta F$ is the change of the free energy. When the internal energy remains a constant, Eq. (\ref{WF}) reduces to $W \leq T\Delta S$. In the informational perspective, the increase of entropy is equivalent to the loss of information, i.e., $\Delta S=-k\Delta I$, so we can write \cite{parrondo2015,Sagawa2008,Horowitz2010,Sagawa2012}
\begin{equation}\label{WI}
W \leq -kT\Delta I.
\end{equation}
This inequality quantitatively relates information to work: greater system knowledge permits more work extraction. Of course, the upper limit $-kT\Delta I$ may not be reached in a process. To extract work as much as possible, one should design the process properly to reduce the dissipation of information. 

The information-work relation offers a deeper understanding of the Gibbs paradox. Crucially, whether the types of gases are same is not essential in the question, what essential is our knowledge of the particle partitioning. To illustrate this point, consider a special scenario where the gases are identical, but we know exactly which side each particle originally occupied. In this case, we could still extract work by using a special membrane that allows only particles from one side to pass. This kind of special membrane is possible in principle, because in classical statistics, all particles are distinguishable and the membrane can be designed carefully to response to the distinguishable properties of each particle in a desired way. 

\section{Discussion and conclusion}
In this letter, we resolve the Gibbs paradox in a clear and simple way. Our resolution uses no additional assumption except the fundamental equal probability principle in classical ensemble theory. It means that the original version of classical ensemble theory by Gibbs (without $1/N!$ factor)   in fact does not cause the Gibbs paradox. It is somewhat surprising that such an ``orthodox'' resolution is not well acknowledged throughout the long history. In our view, this oversight may be attributed to the lack of an informational perspective. Without this perspective, our resolution is hard to understand. For instance, the additivity of entropy is breaking in our resolution, which seems strange except adopting the informational perspective. 

In history, the informational interpretation of entropy is presented by some authors \cite{Jaynes1957,Tseng2002,Lairez2024}, but has not become the mainstream. One important figure in this line is Jaynes, who did many pioneer works on statistical mechanics from the informational perspective \cite{Jaynes1957,Jaynes1957b,Jaynes1965,Jaynes1980,Jaynes1992}. In Jaynes' statistical mechanics, the basic principle is the maximum entropy principle \cite{Jaynes1957}: under given information, the probability distribution should maximum the entropy. Jaynes also offered profound insights into the Gibbs paradox \cite{Jaynes1992}, which inspired many subsequent studies. However, few of these works fully follow his informational perspective. In the Supplemental Material \cite{SM}, we present a resolution of the Gibbs paradox based on the maximum entropy principle, and show that it is equivalent to the resolution based on the equal probability principle.

Some readers may object that making entropy dependent on our knowledge or information introduces subjectivity into thermodynamics. While a full defense of the informational interpretation of entropy lies beyond this work (see \cite{Jaynes1957,Lairez2024} for more discussions), we emphasize that information has demonstrably objective physical consequences -- most notably in determining the amount of controllable work extractable from a system. We believe the perspective presented in this work signifies a paradigm shift in statistical mechanics, in which information will become a central concept and inspire broad potential applications.

\appendix
\section*{Appendix A: Composition law}
\setcounter{equation}{0}
\renewcommand{\theequation}{A\arabic{equation}}
Shannon's definition of information entropy is the unique definition satisfies three basic rules \cite{Shannon1948}. One of the rules is the composition law. Here we give a brief introduction to it.

Consider a variable $x$ with taking values $x_1,x_2,...,x_n$ with corresponding probabilities $p_1,p_2,...,p_n$. The information entropy of this distribution is defined as 
\begin{equation}\label{dis}
S(p_1,p_2,...,p_n)=-\sum_i p_i\ln p_i,
\end{equation}
which measures the uncertainty of the value of $x$. Note here we use natural logarithm, which differs from the base-2 definition by a constant factor.

Now suppose we pack the values $x_1,x_2,...,x_n$ into $r$ disjoint subsets(groups):
\begin{equation}
\begin{aligned}
&g_1=\{x_1,x_2,...,x_{m_1}\},\\
&g_2=\{x_{m_1+1},x_{m_1+2},...,x_{m_1+m_2}\},\\
&........\\
&g_r=\{x_{m_{r-1}+1},x_{m_{r-1}+2},...,x_{m_{r-1}+m_r}\},\\
\end{aligned}
\end{equation}
with 
\begin{equation}
m_1+m_2+...+m_r=n.
\end{equation}
We denote the probabilities for $x$ taking values in these $r$ groups as $w_1,w_2,...,w_r$ respectively, then the information entropy $S(p_1,...,p_n)$ can be decomposed as
\begin{equation}\label{deco2}
\begin{aligned}
S(p_1,...,p_n)=&S(w_1,...,w_r)+w_1S(p_1/w_1,...,p_{m_1}/w_1)\\
&+...+w_rS(p_{m_{r-1}+1}/w_r,...,p_{m_r}/w_r).
\end{aligned}
\end{equation}
The first term $S(w_1,...,w_r)$ measures the uncertainty at the level of groups, and $S(p_{m_{i-1}+1}/w_i,...,p_{m_i}/w_i)$ measures the uncertainty within group $g_i$. Eq. (\ref{deco2}) is called the composition law of information entropy.

The above definition can be generalized to continuous probability distributions. Consider a continuous variable (or set of variables) $x$ distributed over a region $R$ with probability density $\rho(x)$. The corresponding information entropy is:
\begin{equation}
S[\rho(x)]=-\int_R \rho(x)\ln \rho(x)dx.
\end{equation}

Now we divide the region $R$ into $r$ disjoint subregions $R_1,R_2,...,R_r$:
\begin{equation}
\begin{aligned}
&R=R_1\cup R_2 \cup ... \cup R_r, \\
& R_i \cap R_j=\emptyset, \mathrm{for} \ \ i\not= j.
\end{aligned}
\end{equation}
We denote the corresponding probabilities for $x$ taking values in these $r$ regions as $w_1,w_2,...,w_r$ respectively, then the information entropy $S[\rho(x)]$ can be decomposed as
\begin{equation}
\begin{aligned}
S[\rho(x)]=&S(w_1,w_2,...,w_r)+\\
&\ \ w_1S_1[\rho(x)/w_1]+...+w_rS_r[\rho(x)/w_r].
\end{aligned}
\end{equation}
where $S(w_1,w_2,...,w_r)$ is the entropy for the discrete distribution$\{w_i\}$ (see definition in Eq. (\ref{dis})). It measures the uncertainty at the group level. $S_i$ is defined as
\begin{equation}
S_i[\rho(x)/w_i]=-\int_{R_i} \frac{\rho(x)}{w_i}\ln \frac{\rho(x)}{w_i}dx,
\end{equation}
which measures the uncertainty within the subregion $R_i$.

\section*{Appendix B: Derivation of Eq. (16)}
\setcounter{equation}{0}
\renewcommand{\theequation}{B\arabic{equation}}
We now derive Eq.~(16) from the composition law of information entropy. Consider the setup where a box is partitioned by a wall into two equal halves, each containing $N$ particles of the same type. As we have discussed in the main text, the possible phase space region $R$ is divided into $W_d=(2N)!/(N!)^2$ disconnected subregions $R_1,R_2,...,R_{W_d}$, each corresponding to a distinct way of assigning $N$ particles to each side. Due to the permutation symmetry, the probability distribution $\rho(p,q)$ restricted to each subregion is identical, therefore, the weight associated with each subregion becomes:
\begin{equation}
w_1=w_2=...=w_r=\frac{1}{W_d},
\end{equation}
where $w_i= \int_{R_i} \rho(p,q)dpdq$.
Using the composition law of information entropy, the total entropy $S[\rho(p,q)]$ can be decomposed as:
\begin{equation}\label{SSd}
\begin{aligned}
S[\rho(p,q)]&=S(w_1,w_2,...,w_{W_d})+\sum_{i=1}^{W_d} w_i S_i[\rho(p,q)/w_i]\\
&=k\ln W_d+ \sum_{i=1}^{W_d} \frac{1}{W_d} S_i[\rho(p,q)/w_i],
\end{aligned}
\end{equation}
where the group-level entropy is: 
\begin{equation}
S_d=S(w_1,w_2,...,w_{W_d})=k\ln W_d,
\end{equation}
and
\begin{equation}\label{Si}
S_i[\rho(p,q)/w_i]=-k\int_{R_i} \frac{\rho(p,q)}{W_d}\ln \frac{\rho(p,q)}{W_d}dpdq.
\end{equation}
Again, due to the permutation symmetry, all $S_i[\rho(p,q)/w_i]$ are equal. We denote this value as $S_0$, then Eq.(\ref{SSd}) reduces to 
\begin{equation}
S[\rho(p,q)]=S_d+S_0.
\end{equation}
Here, $S_d$ quantifies our ignorance about the particle partition, and $S_0$ measures our uncertainty within each subregion in phase space, i.e., our ignorance about the state of the system when the particle partition is fixed. 

Now let us decompose $S_0$ further. When the way of particle partition is fixed, the left side is independent of the right side, then the distribution $\rho(p,q)/W_d$ within the given subregion can be written as 
\begin{equation}
\frac{\rho(p,q)}{W_d}=\rho_{A}(p_A,q_A)\rho_{B}(p_B,q_B).
\end{equation} 
where $\rho_A(p_A,q_A)$ and $\rho_{B}(p_B,q_B)$ are distributions for left side and right side respectively. Plug this decomposition into Eq. (\ref{Si}), we obtain
\begin{equation}
S_0=S_A+S_B,
\end{equation}
where 
\begin{equation}
\begin{aligned}
&S_A=-k\int \rho_A(p_A,q_A)\ln\rho_A(p_A,q_A)dp_Adq_A,\\
&S_B=-k\int \rho_B(p_B,q_B)\ln\rho_B(p_B,q_B)dp_Bdq_B.\\
\end{aligned}
\end{equation}
Combining everything, the total entropy becomes: 
\begin{equation}\label{de}
S=S_d+S_A+S_B,
\end{equation}
which is exactly Eq. (16) in the main text.


The derivation here reveals that Eq. (\ref{deco}) is a natural consequence of the composition law of information entropy. Except this basic law and the symmetry principle, we make no reference to details of the system in the derivation. Thus, the decomposition in Eq. (\ref{de}) is also valid for more general situations, such as situations with interactions.

\section*{Acknowledgements}
The author acknowledges Chen Zhang for valuable comments on an earlier version of the manuscript. 
\bibliography{ref}

\newpage

\AtEndDocument{\includepdf[pages=1]{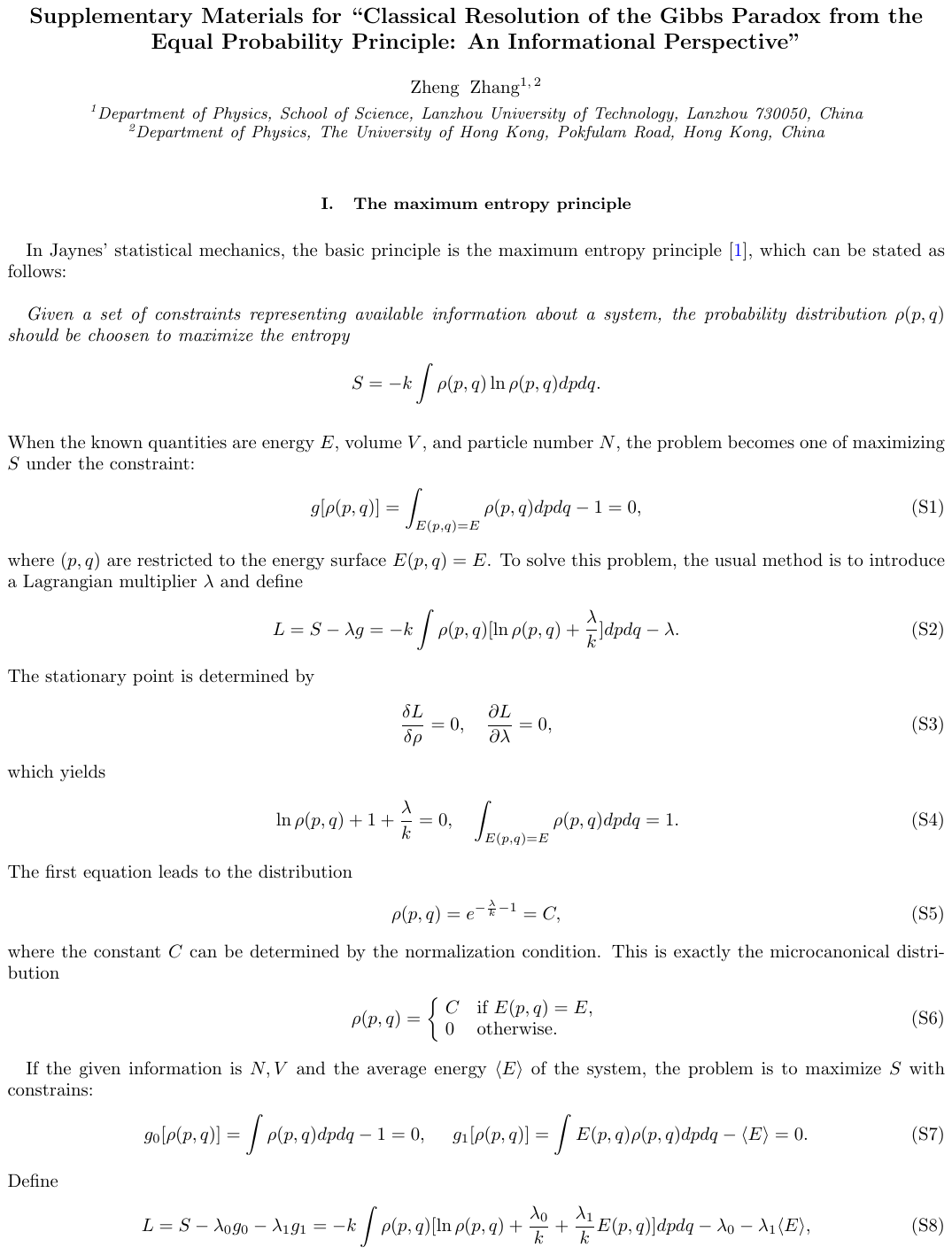}}

\AtEndDocument{\includepdf[pages=2]{supp.pdf}}
\AtEndDocument{

\section*{Supplementary material}

# Supplementary Materials for "Classical Resolution of the Gibbs Paradox from the Equal Probability Principle: An Informational Perspective"

Zheng Zhang[1,2]

[1]Department of Physics, School of Science, Lanzhou University of Technology, Lanzhou 730050, China
[2]Department of Physics, The University of Hong Kong, Pokfulam Road, Hong Kong, China

## I. The maximum entropy principle

In Jaynes' statistical mechanics, the basic principle is the maximum entropy principle [1], which can be stated as follows:

*Given a set of constraints representing available information about a system, the probability distribution $\rho(p,q)$ should be choosen to maximize the entropy*

$$S = -k \int \rho(p,q) \ln \rho(p,q) dp dq.$$

When the known quantities are energy $E$, volume $V$, and particle number $N$, the problem becomes one of maximizing $S$ under the constraint:

$$g[\rho(p,q)] = \int_{E(p,q)=E} \rho(p,q) dp dq - 1 = 0, \tag{S1}$$

where $(p,q)$ are restricted to the energy surface $E(p,q) = E$. To solve this problem, the usual method is to introduce a Lagrangian multiplier $\lambda$ and define

$$L = S - \lambda g = -k \int \rho(p,q)[\ln \rho(p,q) + \frac{\lambda}{k}] dp dq - \lambda. \tag{S2}$$

The stationary point is determined by

$$\frac{\delta L}{\delta \rho} = 0, \quad \frac{\partial L}{\partial \lambda} = 0, \tag{S3}$$

which yields

$$\ln \rho(p,q) + 1 + \frac{\lambda}{k} = 0, \quad \int_{E(p,q)=E} \rho(p,q) dp dq = 1. \tag{S4}$$

The first equation leads to the distribution

$$\rho(p,q) = e^{-\frac{\lambda}{k}-1} = C, \tag{S5}$$

where the constant $C$ can be determined by the normalization condition. This is exactly the microcanonical distribution

$$\rho(p,q) = \begin{cases} C & \text{if } E(p,q) = E, \\ 0 & \text{otherwise.} \end{cases} \tag{S6}$$

If the given information is $N, V$ and the average energy $\langle E \rangle$ of the system, the problem is to maximize $S$ with constrains:

$$g_0[\rho(p,q)] = \int \rho(p,q) dp dq - 1 = 0, \quad g_1[\rho(p,q)] = \int E(p,q)\rho(p,q) dp dq - \langle E \rangle = 0. \tag{S7}$$

Define

$$L = S - \lambda_0 g_0 - \lambda_1 g_1 = -k \int \rho(p,q)[\ln \rho(p,q) + \frac{\lambda_0}{k} + \frac{\lambda_1}{k} E(p,q)] dp dq - \lambda_0 - \lambda_1 \langle E \rangle, \tag{S8}$$

The stationary point is determined by

$$\frac{\delta L}{\delta \rho} = 0, \quad \frac{\partial L}{\partial \lambda_0} = 0, \quad \frac{\partial L}{\partial \lambda_1} = 0, \tag{S9}$$

which yields

$$\ln \rho(p,q) + 1 + \frac{\lambda_0}{k} + \frac{\lambda_1}{k} E(p,q) = 0, \quad \int \rho(p,q) dp dq = 1, \quad \int E(p,q) \rho(p,q) dp dq = \langle E \rangle. \tag{S10}$$

The first equation gives

$$\rho(p,q) = \frac{e^{-\frac{\lambda_1}{k} E(p,q)}}{e^{1+\frac{\lambda_0}{k}}} = \frac{e^{-\beta E(p,q)}}{Z}, \tag{S11}$$

where $Z$ and $\beta$ are determined by the constrains, which can be equivalently written as

$$Z = \int e^{-\beta E(p,q)} dp dq, \quad \langle E \rangle = -\frac{\partial \ln Z}{\partial \beta}. \tag{S12}$$

Eq. (S11) is just the canonical distribution, and $\beta = -\frac{1}{kT}$.

Finally, when the given information includes $V$, $\langle E \rangle$ and the average particle number $\langle N \rangle$, the problem is to maximize $S$ with constrains:

$$g_0[\rho(p,q)] = \int \rho(p,q) dp dq - 1 = 0, \quad g_1[\rho(p,q)] = \int E(p,q) \rho(p,q) dp dq - \langle E \rangle = 0,$$
$$g_2[\rho(p,q)] = \sum_{N=0}^{\infty} N \rho_N(p,q) - \langle N \rangle = 0, \tag{S13}$$

where $\rho_N(p,q)$ is the distribution $\rho(p,q)$ restricted to the $N$-particle subspace. Define

$$L = S - \lambda_0 g_0 - \lambda_1 g_1 - \lambda_2 g_2$$
$$= -k \sum_{N=0}^{\infty} \int \rho_N(p,q) [\ln \rho_N(p,q) + \frac{\lambda_0}{k} + \frac{\lambda_1}{k} E_N(p,q) + \frac{\lambda_2}{k} N] dp dq - \lambda_0 - \lambda_1 \langle E \rangle - \lambda_2 \langle N \rangle, \tag{S14}$$

The stationary point satisfies

$$\frac{\delta L}{\delta \rho} = 0, \quad \frac{\partial L}{\partial \lambda_0} = 0, \quad \frac{\partial L}{\partial \lambda_1} = 0, \quad \frac{\partial L}{\partial \lambda_2} = 0, \tag{S15}$$

which yields

$$\ln \rho_N(p,q) + 1 + \frac{\lambda_0}{k} + \frac{\lambda_1}{k} E_N(p,q) + \frac{\lambda_2}{k} N = 0,$$
$$\int \rho(p,q) dp dq = 1, \quad \sum_{N=0}^{\infty} \int E(p,q) \rho_N(p,q) dp dq = \langle E \rangle, \quad \sum_{N=0}^{\infty} N \rho_N(p,q) = \langle N \rangle. \tag{S16}$$

The first equation gives

$$\rho_N(p,q) = \frac{e^{-\frac{\lambda_1}{k} E(p,q) - \frac{\lambda_2}{k} N}}{e^{1+\frac{\lambda_0}{k}}} = \frac{e^{-\beta E(p,q) - \alpha N}}{\Xi}, \tag{S17}$$

where $\Xi$, $\beta$ and $\alpha$ are determined by the constrains, which can be equivalently written as

$$\Xi = \sum_{N=0}^{\infty} \int e^{-\beta E_N(p,q) - \alpha N} dp dq, \quad \langle E \rangle = -\frac{\partial \ln \Xi}{\partial \beta}, \quad \langle N \rangle = -\frac{\partial \ln \Xi}{\partial \alpha}. \tag{S18}$$

Eq. (S17) is just the grandcanonical distribution, and $\beta = -\frac{1}{kT}$, $\alpha = -\frac{\mu}{kT}$.

## II. Resolution of the Gibbs paradox

We now apply the maximum entropy principle to resolve the Gibbs paradox, focusing on the case where the gases in both sides are of the same type.

Suppose the box is divided into two equal halves by an impermeable wall. The total volume is $2V$, and the total number of particles is $2N$, with exactly $N$ particles on each side. This additional constraint—on the particle number in each subvolume—represents extra information compared to the standard canonical ensemble. This additional information can regarded as a constraint on the phase space, i.e.,

$$
\begin{aligned}
0 < q_{\sigma(i),x} < L, & \quad \text{for } i = 1, 2, ..., N, \\
L < q_{\sigma(i),x} < 2L, & \quad \text{for } i = N+1, N+2, ..., 2N.
\end{aligned}
\tag{S19}
$$

where $\sigma(i)$ is a permutation on the indices $1, 2, ..., 2N$. The possible phase space is given by running over all permutations. Except for this extral constraint on the phase space, the problem is the same as the case of canonical ensemble, i.e., maximizing the entropy with the constraints

$$
g_0[\rho(p,q)] = \int \rho(p,q) dp dq - 1 = 0, \quad g_1[\rho(p,q)] = \int E(p,q) \rho(p,q) dp dq - \langle E \rangle = 0.
\tag{S20}
$$

Repeat the procedures in last section, we get the distribution

$$
\rho(p,q) = \frac{e^{-\beta E(p,q)}}{Z},
\tag{S21}
$$

with $Z$ and $\beta$ determined by

$$
Z = \int e^{-\beta E(p,q)} dp dq, \quad \langle E \rangle = -\frac{\partial \ln Z}{\partial \beta}.
\tag{S22}
$$

Note that the integration over phase space now includes only the allowed regions consistent with the particle-number constraint. This result is exactly that we obtained the main text, and then the Gibbs paradox can be resolved in a same manner.

---

}
\end{document}